\documentclass[12pt]{article}
\usepackage{graphicx}

\begin{document}

\begin{center}
{\bf {APPLICATION OF ELASTIC MID$\,$-IR$\,$-LASER$\,$-LIGHT SCATTERING FOR
NON-DESTRUCTIVE INSPECTION IN MICROELECTRONICS}}
\end{center}

\vspace*{0mm}

{\flushleft {VICTOR P. KALINUSHKIN$^{^{\ast}}$, VLADIMIR A.
YURYEV$^{^{\ast}}$, OLEG V. ASTAFIEV$^{^{\ast}}$,}}\\
ALEXANDER N. BUZYNIN$^{^{\ast}}$, AND NIKOLAY I. BLETSKAN$^{^{\ast \ast}}$ \\
$^{^{\ast}}$General Physics Institute of RAS, 38  Vavilov  Street,  Moscow,
117942, Russia\\
$^{^{\ast \ast}}$Research and Production Association ELMA,
Zelenograd, Moscow, 103482, Russia\\
\vspace*{0mm}

{\flushleft {\bf {ABSTRACT}}}\\

Some possible applications of the low-angle mid-IR-light scattering
technique and some recently developed on its basis methods for
non-destructive inspection and investigation of semiconductor materials
and structures are discussed in the paper. The conclusion is made
that the techniques in question might be very useful for solving
a large number of problems regarding defect investigations
and quality monitoring both in research laboratories
and the industry of microelectronics.\\

{\flushleft {\bf {INTRODUCTION}}}\\

Over 15 years, the method of low-angle mid-IR-light scattering
(LALS) have been actively used by us for investigation of the large-scale
electrically-active defect accumulations (LSDAs) in semiconductors
(see e.g. Refs.[1--15] and references cited therein). We shall consider
below some possible
applications of LALS and techniques recently developed on its basis
for solving some specific problems of material and structure testing in
microelectronics.\\

{\flushleft {\bf {BRIEF DESCRIPTION}}}\\

For the beginning let us briefly remind the basic principles of
LALS. This method is founded on elastic scattering of IR light by
inhomogeneities of semiconductor crystals$\,$---$\,$like the method of laser
tomography,---$\,$but in LALS, as distinct from the laser tomography, the
scattering at relatively low angles is registered$\,$---$\,$from
about 2$^{\circ}$ to
less than 15$^{\circ}$  in crystal$\,$---$\,$and light with large
wavelength is used as a
probe emission$\,$---$\,$routinely the radiation of CO$_2$- or
CO-lasers with the
wavelength of 10.6 $\mu$m and 5.4 $\mu$m, respectively, is applied.
Application of mid-IR-light
makes the technique sensitive to the presence of domains with
enhanced concentration of free carrier or changed conductance type
(FCAs) and measuring in the above interval of angles allows one to
observe defects with the sizes from several microns to several dozens
of microns.${\sl {^{1,2,9,11}}}$ (Writing FCAs we mean both
manmade domains with changed carrier concentration or conductance
type$\,$---$\,$e.g. doped domains of semiconductor structures$\,$---$\,$and
natural LSDAs which always are FCAs.)

A number of procedures has been developed, which enable the
distinguishing of the scattering by FCAs from that by different
defects. These procedures consist in the investigation of the light
scattering with different wavelengths or measuring the dependencies of
the light scattering intensity on sample temperature.${\sl {^{9,12,14,15}}}$
In addition, the latter procedure and the investigation of the influence
of photoexcitation on light scattering intensity allows one to
determine thermal and optical activation energies of point centers in
LSDAs.${\sl {^{9,12}}}$ Moreover, LALS allows one to investigate large-scale
recombination-active defects (LSRDs) and large-scale gluing centers
(LSGCs) discriminating between those in near-surface
layers$\,$---$\,$including
epilayers$\,$---$\,$and those in substrate bulk: the former are studied by use
of the surface optical excitation,${\sl {^{16}}}$ while the latter
are observed
using the volume photoexcitation.${\sl {^{1,8,10}}}$ (Typical examples
of LSRDs
are such defects as grain boundaries, dislocations, swirls,
precipitates, their clusters and aggregations of recombination point
centers.)

\begin{figure}
  \begin{center}
 \includegraphics[scale=1]{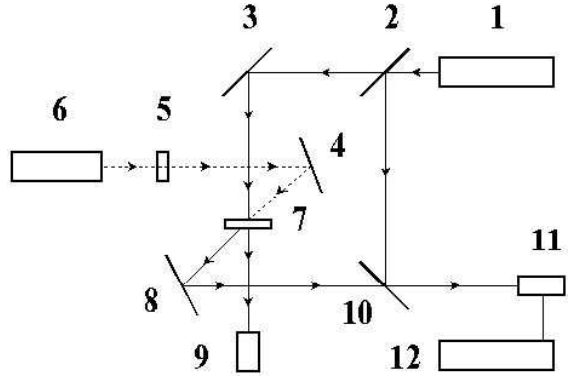}
 \end{center}
   \caption{Optical diagram of angle-resolved LALS: (1)
mid-IR-laser;  (2,10)
semitransparent mirrors; (3,4) mirrors; (5) filters; (6) exciting laser
(used in LALS with photoexcitation); (7)  sample;  (8)  movable  mirror;
(9,11) photoreceivers; (12) computer.}
\end{figure}

\begin{figure}
  \begin{center}
 \includegraphics[scale=1]{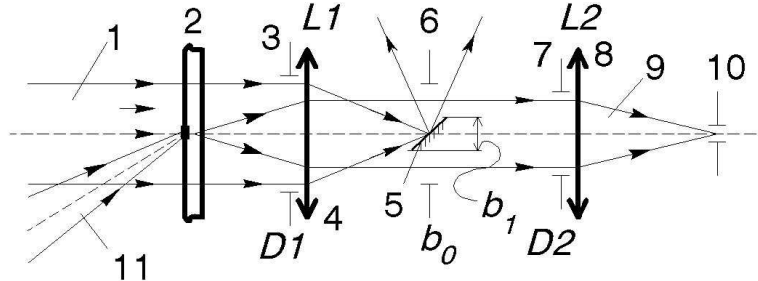}
 \end{center}
   \caption{Optical diagram of the SLALS microscope:
(1) mid-IR  probe  wave;
(2) sample; (3,6,7) diaphragms;  (4,8)  lenses;  (5)  mirror  or  opaque
screen; (9) scattered wave; (10) IR photodetector; (11)  exciting  light
beam (used in OLALS).}
\end{figure}

Presently, the two following schemes of LALS are developed. One of
them$\,$---$\,$the conventional LALS or LALS with angular
resolution$\,$---$\,$registers
light scattered by all defects which are situated within the
probe beam,${\sl {^{2,10,12}}}$ see Fig.1. In the other$\,$---$\,$in
scanning LALS
microscopy or SLALS, which is a kind of the scanning laser
microscopy,---$\,$every singular defects are visualized,${\sl {^{17-20}}}$ see
Fig.2. By combining these two schemes, one can determine concentration
of defects, their spatial distribution, and hence, one can estimate
the value of deviation of their dielectric constants from those of
crystal bulk outside them ${\Delta \epsilon}$ (in the case of
FCAs$\,$---$\,$and, as mentioned
above, LSDAs always are FCAs,---$\,$the concentrations of free carriers in
them, ${\Delta n}$, can be evaluated).${\sl {^{9,10,14,15}}}$

   The LALS technique is of high sensitivity, it allows one to observe
defects with the variation of dielectric constant ${\Delta \epsilon}$
down to 10${\rm {^{-5}}}$--10${\rm {^{-4}}}$$\,$---$\,$i.e. with ${\Delta n}$
down to 10${\rm {^{13}}}$ cm${\rm {^{-3}}}$. LALS is non-destructive and
contactless, it has no limitations on the tested wafer diameter.\\

{\flushleft {\bf {INDUSTRIAL AND LABORATORY APPLICATIONS}}}\\

Let us dwell on some possible specific applications of LALS as the
checking technique in the field of the industrial microelectronics.\\

{\flushleft {\underline {Inspection of semiconductor wafer homogeneity}}}\\

As mentioned above,
LALS enables the observation of LSDAs with the sizes from several $\mu$m to
several tens $\mu$m with point detect concentration in each of them down
to 10${\rm {^{13}}}$  cm${\rm {^{-3}}}$. This method permits the
investigation of the LSDA
composition and the influence of various thermal treatments and
operations of an industrial technological cycle on them. Wafer mapping
is possible by means of SLALS. Incoming control with posterior
utilization of substrates in the production process and technological step
checking by using free chips are also possible.

   These techniques are well developed now and a prototype of the
instrument is available.

   Fig.3 demonstrates the images of LSDAs in different bulk semiconductors
obtained with the SLALS microscope (1$\times$1 mm${\rm {^{2}}}$  areas are
presented). The
microphotographs of LEC undoped InP $(a)$, LEC InP:Fe $(b)$, LEC SI GaAs
annealed at 900$^{\circ}$C in a sealed quartz ampule when
produced $(c)$, CZ Si:B with high $(d)$ and low $(e)$ epd,
and CZ Si:B coated with 1200 {\AA} thick SiO${\rm {_{2}}}$  layer $(f)$
are presented. White spots are the images of LSDAs. One can find more
details on these pictures in Refs.[18-21].

\begin{figure}
 \begin{center}
 \includegraphics[scale=1]{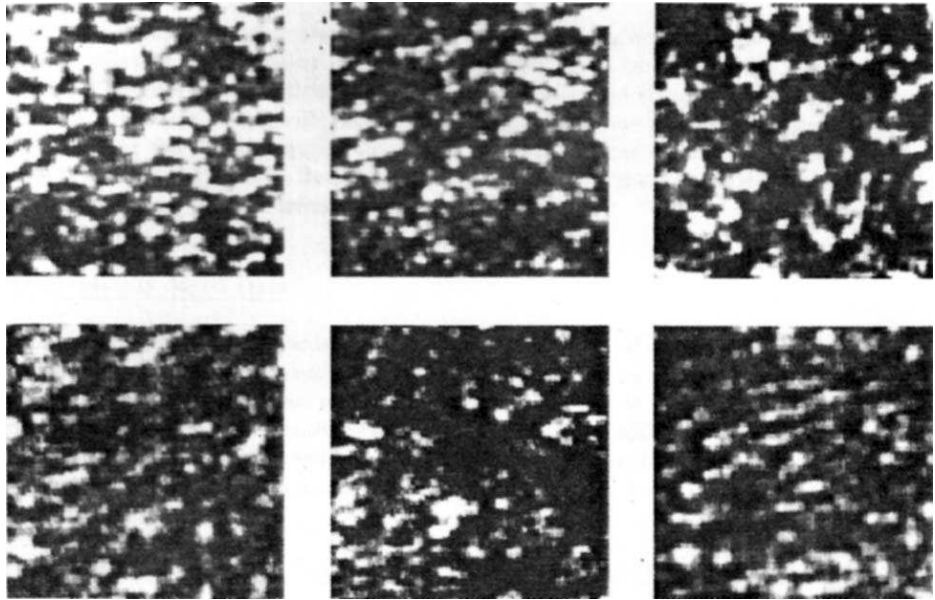}
 \end{center}
   \caption{SLALS images of LSDAs in semiconducting  wafers;
the  first  row
(left to right): undoped LEC InP  $(a)$;  LEC  InP:Fe  $(b)$;  LEC  SI  GaAs
annealed at 900$^{\circ}$C $(c)$; the second row (left to right): CZ Si:B,
$\rho \,$=12 $\Omega \,$cm,
high epd $(d)$; CZ Si:B,  $\rho \,$=12 $\Omega \,$cm,  low  epd  $(e)$;
CZ  Si:B (different
establishment) under 1200 {\AA} thick SiO${\rm {_{2}}}$
layer $(f)$; 1$\times$1
mm${\rm {^{2}}}$ , $\lambda {\sf {_{sc}}}$= 10.6 $\mu$m.
Amplification factors are equal in the following pairs  of  pictures:
$(a)$  and
$(b)$, $(d)$ and $(e)$. White spots in  the  photographs  are  the  images  of
LSDAs.}

\end{figure}

   The disadvantage of LALS (and SLALS) is its inability to
discriminate between LSDAs situated in crystal bulk and ones located
in near-surface layer. To remove this shortcoming, LALS tomography
with longitudinal resolution down to 10--20 $\mu$m is now under development
on the basis of SLALS. The solution of the problem does not seem to meet any
difficulties. Its successful solving would enable the
testing of homogeneity of "working" near-surface layers of wafers and
epitaxial layers. Note that layer inspection is possible even if it is
under coating or under other layer$\,$---$\,$up to the stage of metallization.
For instance, the homogeneity of a silicon wafer  under
oxide layer may be checked$\,$---$\,$Fig.3$(d) \,$---$\,$as well as
its near-surface layer .\\

{\flushleft {\underline {Inspection of presence of LARDs in near-surface,
near-interface and epitaxial layers}}}\\

The methods of optical beam induced LALS (OLALS) and LALS with
optical pumping,${\sl {^{16}}}$ which may be used for such inspections, are
the optical analogs of such well-known methods as EBIC and OBIC,
yet they require neither Shottky barrier or $p-n$ junction nor
complicated sample preparation. Like in the above case, a
significant advantage of these techniques is their ability to test
multilayer structures including layers covered with other layers. Also
there are no limitations on sample size and resistivity in LALS. So
wafer mapping, all-round incoming and step control with subsequent
utilization of substrates in production cycle are possible.

   Control of LSRDs in the substrate volume, including tomography, is
also possible, but this is likely of interest for the production of
$\gamma$-ray detectors, whose resolution is determined by LSDAs and
LSRDs,${\sl {^{10}}}$ various nuclear-ray counters, volume photodetectors,
{\em {etc}}.

   As of now, a cycle of experiments has been carried out, which have
shown the possibilities of these methods.${\sl {^{8,10,16,19,20}}}$

\begin{figure}
 \begin{center}
 \includegraphics[scale=1]{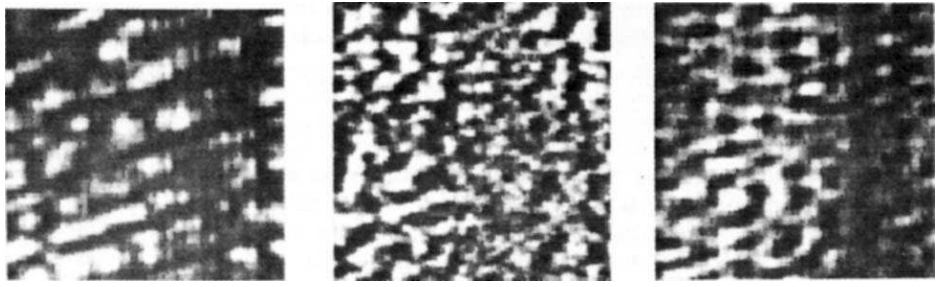}
 \end{center}
  \caption{OLALS microphotographs of Si wafers, defects
in near-surface layers (left to right): FZ Si:P,
chemico-dinamic  polishing  $(a)$;  FZ  Si:P,
mechanical polishing $(b)$; CZ Si:B under 1200 {\AA} thick SiO${\rm {_{2}}}$
layer $(c) \,$---$\,$the same region as that given in Fig.3$(f)$; 1$\times$1
mm${\rm {^{2}}}$ , $\lambda {\sf {_{sc}}}$= 10.6 $\mu$m,
$\lambda {\sf {_{ex}}}$=
633 nm. Amplification factor in picture $(b)$ is by 100  times  greater
than  in
picture $(a)$, and that in picture $(c)$ is by 10 times greater than in
picture
$(a)$.  The  brighter  image,  the  longer  the  non-equilibrium   carrier
effective lifetime is. The dark areas are the images of defective regions.}
\end{figure}

   Fig.4 demonstrates the microphotographs of near-surface regions of
Si wafers subjected to different polishing and oxidation procedures
(1$\times$1 mm${\rm {^{2}}}$  areas are presented). The images of FZ Si:P
wafers after
chemico-dinamic $(a)$ and mechanical $(b)$ polishing procedures, and the
image of CZ Si:B wafer coated with 1200 {\AA} thick SiO${\rm {_{2}}}$
layer $(c)$ are given in this figure. The darker image, the shorter
non-equilibrium  carrier effective lifetime is.
The focused 633-nm-wavelength radiation
of 55-mW He-Ne laser was used in this experiment for electron-hope pairs
photoexcitation in near-surface layers (see Fig.2).\\

{\flushleft {\underline {Testing of specially doped areas}}}\\

LALS may be also used for
the inspection of such parameters of specially doped domains of semiconductor
structures as their sizes, concentration of free carrier in them and
surface resistance. The inspection of these parameters is also
possible even after different layers are grown and coatings are given
(until metallized). The domains with the sizes greater than 1 $\mu$m and
the variation of free carrier concentration greater than 10${\rm {^{13}}}$
cm${\rm {^{-3}}}$ can be tested.
Nowadays, the development of a prototype of such instrument is being
in the final stage.\\

{\flushleft {\underline {Inspection of gettering process efficiency}}}\\

We would like to specially emphasize that the above techniques might be
very useful for the inspection of the gettering processes
efficiency.${\sl {^{22}}}$

   The presence of the gettering precipitates at the internal gettering
process can be checked by OLALS or LALS with quasi-bulk
photoexcitation. The presence and parameters of impurity atmospheres
around the gettering precipitates can be checked by the conventional
(angle-resolved) LALS and SLALS microscopy. The "working layer" may
be tested by OLALS or LALS with surface photoexcitation.

   The inspection of efficiency of the external gettering and gettering
by implanted
domains  is also possible by the LALS-based techniques. The
procedures proposed for these inspections are analogous to those
described above for the internal gettering process.\\

{\flushleft {\bf {CONCLUSION}}}\\

   So we can conclude that the LALS-based techniques might by a very
effective non-destructive tool for solving a wide class of problems of
materials and structures testing in modern microelectronics, which
might be used both in laboratories and directly in the production
cycle. We discussed only several most obvious possible
applications of these techniques in this paper. We are sure, however,
that they might find a great number of additional applications and be
useful in many branches of microelectronics science and industry.\\

{\flushleft {\bf {REFERENCES}}\\\

1. V.V. Voronkov, T.M. Murina, G.I. Voronkova}
{\sl {et al.}}, Fiz. Tverd. Tela {\bf {20}}, 1365 (1978)

[Sov. Phys. Solid State {\bf {20}} (5), 1365 (1978)].\\
2. V.V. Voronkov, G.I. Voronkova, B.V. Zubov {\sl {et al.}},
Fiz. Tverd. Tela {\bf {23}} (1), 117 (1981)

[Sov. Phys. Solid State, {\bf {23}} (1), 65 (1981)].\\
3. V.V. Voronkov, G.I. Voronkova, V.N. Golovina {\sl {et al.}},
J. Cryst. Growth {\bf {52}}, 939
(1981).\\
4. V.V. Voronkov, G.I. Voronkova, V.P. Kalinushkin {\sl {et al.}},
Fiz. Tekh. Poluprovodn. {\bf {18}}

(12), 2222 (1984)
[Sov. Phys. Semicond. {\bf {18}} (12), 2222 (1984)].\\
5. S.E. Zabolotskii, V.P. Kalinushkin, T.M. Murina {\sl {et al.}},
Phys. Stat. Sol.(a) {\bf {88}}, 539

(1985).\\
6. N.V. Veselovskaya, V.V. Voronkov, G.I. Voronkova {\sl {et al.}},
Fiz. Tverd. Tela {\bf {27}} (5), 1331

(1985)
[Sov. Phys. Solid State {\bf {27}} (5), 1331 (1985)].\\
7. A.V. Voronkova, V.P. Kalinushkin, T.M. Murina, and N.S. Sysoyeva
Fiz. Tekh.

Poluprovodn. {\bf {19}} (10), 1902 (1985)
[Sov. Phys. Semicond. {\bf {19}} (10), 1902 (1985)].      \\
8. V.P. Kalinushkin, D.I. Murin, T.M. Murina {\sl {et al.}},
Microelectronica {\bf {15}} (6), 523 (1986)

[Sov. Phys. Microelectronics {\bf {15}} (6), 523 (1986)].\\
9. S.E.Zabolotskii, V.P.Kalinushkin, D.I.Murin {\sl {et al.}},
Fiz. Tekh. Poluprovodn. {\bf {21}} (8),

1364 (1987)
[Sov. Phys. Semicond. {\bf {21}} (8), 1364 (1987)].       \\
10. Victor P. Kalinushkin, in
{\em {Proc. Inst. Gen. Phys. Acad. Sci. USSR, Vol.4, Laser

Methods of Defect
Investigations in Semiconductors and Dielectrics}}, edited by

A.A. Manenkov
(Nova, New York, 1988) pp. 1--75.                           \\
11. A.V. Batunina, V.V. Voronkov, G.I. Voronkova {\sl {et al.}},
Fiz. Tekh. Poluprovodn. {\bf {22}} (7),

1308 (1988)
[Sov. Phys. Semicond. {\bf {22}} (7), 1308 (1988)].   \\
12. V.V. Voronkov, V.P. Kalinushkin, D.I. Murin {\sl {et al.}}
J. Cryst. Growth {\bf{103}}, 126--130

(1990).           \\
13. A.N. Buzynin, S.E. Zabolotskii, V.P. Kalinushkin {\sl {et al.}},
Fiz. Tekh. Poluprovodn. {\bf {24}}

(2), 264 (1990)
[Sov. Phys. Semicond. {\bf {24}} (2), 264 (1990)].   \\
14. V.P. Kalinushkin, V.A. Yuryev, and D.I. Murin,
Fiz. Tekh. Poluprovodn. {\bf {25}}, 798

(1991)
[Sov. Phys. Semicond. {\bf {25}} (5), 798 (1991)].\\
15. V.P. Kalinushkin, V.A. Yuryev, D.I. Murin, and M.G. Ploppa, Semicond.
Sci.

Technol. {\bf 7}, A255--A262 (1992).  \\
16. V.P. Kalinushkin, D.I. Murin, V.A. Yuryev {\sl {et al.}},
in {\em {Second International
Symposium

on Advanced Laser Technologies}}, edited by V. Pustovoy and
M. Jel\'{\i}nek,
Proc. SPIE

{\bf {2332}}, 146--153 (1994).\\
17. O.V. Astafiev, V.P. Kalinushkin, and V.A. Yuryev,
in {\em {Second International Symposium

on Advanced Laser Technologies}}, edited by V. Pustovoy and
M. Jel\'{\i}nek,
Proc. SPIE

{\bf {2332}}, 138--145 (1994).\\
18. O.V. Astafiev, V.P. Kalinushkin, and V.A. Yuryev,
Mater. Sci. Eng. B (submitted for

publication).\\
19. V.P. Kalinushkin, V.A. Yuryev, and O.V. Astafiev,
presented at the First International

Conference on Materials for Microelectronics, Barcelona,
Spain, October 18--23, 1994

(Mater. Sci. Technol, 1995) (submitted for
publication).\\
20. O.V. Astafiev, V.P. Kalinushkin, and  V.A. Yuryev,
presented at the Ninth International

Conference on Microscopy of
Semiconducting Materials, Oxford,
UK, March 20--23, 1995

(IOP Conf. Ser.)  (submitted for publication).\\
21. Vladimir A. Yuryev and Victor P. Kalinushkin, Mater. Sci. Eng. B
(in the press).\\
22. V.P. Kalinushkin, A.N. Buzynin, D.I. Murin, V.A. Yuryev,
O.V. Astafiev, and

A.I.Buvaltsev, presented at the First International
Conference on Materials for

Microelectronics, Barcelona, Spain, October 18--23, 1994 (unpublished).

\end{document}